\documentclass[prb,twocolumn,floatfix,notitlepage,superscriptaddress,longtable]{revtex4-2}
\usepackage{amsmath}
\usepackage{lipsum} % for some text
\usepackage{verbatim}
\usepackage{epsfig}
\usepackage{subfigure}
\usepackage{graphicx}
\usepackage{amsfonts}
\usepackage[figuresright]{rotating}
\usepackage{amssymb}
\usepackage{amsmath}
\usepackage[normalem]{ulem}
\usepackage{psfrag}
\usepackage{esint} %double integrals
\usepackage{bm}% bold math
\usepackage[colorlinks,linkcolor=blue,anchorcolor=blue,citecolor=blue,urlcolor=blue]{hyperref}
\usepackage[version=4]{mhchem}
\usepackage[svgnames]{xcolor}
\def\be{\begin{equation}} \def\ee{\end{equation}}
\def\bea{\begin{eqnarray}} \def\eea{\end{eqnarray}}

\newcommand{\ket}[1]{| #1 \rangle}
\newcommand{\bra}[1]{\langle #1 |}
\newcommand{\braket}[2]{\langle #1 |#2\rangle}

\def\bpm{\begin{pmatrix}} \def\epm{\end{pmatrix}}
\usepackage{graphicx} % for \resizebox
\usepackage{xcolor}   % for \definecolor and color mixing (arrowred!80!black)
\usepackage{booktabs}
\usepackage{tikz}
\usetikzlibrary{positioning, shapes.geometric, arrows.meta, calc, fit, shadows}
\usetikzlibrary{arrows.meta} % for Stealth arrow tips like -{Stealth}
\usepackage{pgfplots}
\pgfplotsset{compat=1.18}    % or compat=newest

\usepackage{tikz}
\usepackage{graphicx} % Required for \resizebox
\usetikzlibrary{positioning, shapes.geometric, arrows.meta, calc, fit, shadows, backgrounds}

\definecolor{boxblue}{RGB}{218, 232, 252}
\definecolor{boxgreen}{RGB}{213, 232, 212}
\definecolor{boxorange}{RGB}{255, 204, 153}
\definecolor{boxpurple}{RGB}{225, 213, 231}
\definecolor{addyellow}{RGB}{255, 255, 153}
\definecolor{arrowblue}{RGB}{22, 79, 134}
\definecolor{arrowred}{RGB}{178, 34, 34}
\definecolor{panelgray}{RGB}{245, 245, 245}

\newcommand{\edited}[1]{{\color{Red}}}
\definecolor{Qicolor}{RGB}{3, 136, 252}

\makeatletter
\newcommand*{\balancecolsandclearpage}{%
  \close@column@grid
  \clearpage
}
\makeatother

\begin{document}
\title{Neural Operator Quantum State: A Foundation Model for Quantum Dynamics}

\author{Zihao Qi}
\email[Contact author: ]{zq73@cornell.edu}
\affiliation{Department of Physics, Cornell University, Ithaca, NY 14853, USA.}

\author{Christopher Earls}
\affiliation{Center for Applied Mathematics, Cornell University, Ithaca, NY 14853, USA.}

\author{Yang Peng}
\email[Contact author: ]{yang.peng@csun.edu}
\affiliation{Department of Physics and Astronomy, California State University, Northridge, Northridge, California 91330, USA}
\affiliation{Institute of Quantum Information and Matter and Department of Physics,California Institute of Technology, Pasadena, CA 91125, USA}

\date{\today}

\begin{abstract}
Capturing the dynamics of quantum many-body systems under time-dependent driving protocols is a central challenge for numerical simulations. Existing methods such as tensor networks and time-dependent neural quantum states, however, must be re-run for every protocol. In this work, we introduce the Neural Operator Quantum State (NOQS) as a foundation model for quantum dynamics. Rather than solving the Schrödinger equation for individual trajectories, our approach aims to \emph{learn the solution operator} that maps entire driving protocols to time-evolved quantum states. Once trained, the NOQS predicts time evolution under unseen protocols in a single forward pass, requiring no additional optimization. We validate NOQS on the two-dimensional Ising model with time-dependent longitudinal and transverse fields, demonstrating accurate prediction not only for unseen in-distribution protocols, but also for qualitatively different, out-of-distribution functional forms of driving. Further, a single NOQS model can be transferred between different temporal resolutions, and can be efficiently fine-tuned with sparse experimental measurements to improve predictions across all observables at negligible cost. Our work introduces a new paradigm for quantum dynamics simulation and provides a practical computational-experimental interface for driven quantum systems.
\end{abstract}

\maketitle

\section{Introduction}
Characterizing the ground-state properties and dynamical evolution of quantum many-body systems remains a central challenge in physics, as exact representations of the many-body wavefunction become computationally intractable for large systems due to the exponential growth of Hilbert space dimension with the system size.

To overcome this difficulty, a variety of approximate methods have been developed. Matrix product states (MPS) and the density matrix renormalization group (DMRG)~\cite{dmrg_review1, dmrg_review2, dmrg1, dmrg2, mps_dmrg} have achieved remarkable success in one-dimensional systems, but their applicability to higher dimensions is limited. Higher-dimensional tensor-network approaches, such as projected entangled pair states (PEPS)~\cite{PEPS, PEPS2, PEPS3, iPEPS}, can in principle address this limitation, but only at substantially increased computational cost. More generally, in tensor-network methods, the amount of entanglement that can be represented is controlled by the bond dimension, which is itself limited in practice by available computational resources. Consequently, these methods often struggle to capture highly entangled states and are therefore less well suited to quantum dynamics. Traditional variational Monte Carlo (VMC) methods, including time-dependent formulations, also face difficulties in some systems due to the sign problem~\cite{li2019sign}. In addition, their reliance on Metropolis-based Markov-chain sampling leads to correlated samples and can result in non-ergodic behavior when acceptance rates are low.

Deep neural networks, which are capable of approximating arbitrary functions~\cite{NNunivapprox1,NNunivapprox2,NNunivapprox3}, have recently emerged as powerful ansatz for representing quantum wavefunctions in equilibrium and out-of-equilibrium quantum many-body systems~\cite{nqs_review1,nqs_review2, carleo2017,Transformer, nqs-excitedstates, nqs-periodicsystem, nqs-tomography, real-time-evolution-nqs, tNQS_Bohrdt, tNQS_Carleo, nqs_dynamics_recent, nqs_recent, nqs-finetuning, NQS_recent2, nqs-timevolutionscaling, NQS_dynamics_tDVP}. These neural-network quantum states (NQS) are fundamentally appealing because they are not constrained by the area-law entanglement structure characteristic of tensor-network states~\cite{NQSEntanglement}, nor do they, \emph{a priori}, suffer from the sign problem when combined with VMC techniques. Moreover, certain network architectures, such as recurrent neural networks~\cite{RNN, RNN2} and transformers~\cite{vaswani2017attention}, support direct, or \emph{autoregressive}, sampling schemes that generate uncorrelated samples~\cite{sharir2020_autoregressive,RNNauto,barrett2022autoregressive}. With this efficient sampling strategy, autoregressive NQS have been shown to accurately characterize the ground states and time-evolution of various quantum lattice models~\cite{Luo2023,autoregressive-hubbard, luodi_autoregressive_open, barrett2022autoregressive, sharir2020_autoregressive, Donatella_autoregressive_dynamics, tNQS_Bohrdt, autoregressive-quantumnumber}.

Despite the success of autoregressive NQS in modeling ground states, nonequilibrium quantum dynamics has been explored far less extensively within this framework. Typically, a time-dependent NQS (tNQS) is constructed by allowing the parameters of the neural network to depend on time. These parameters are then optimized via the time-dependent variational principle~\cite{tdvp}. It has been shown that tNQS can approximate the time evolution of a quantum state under a given time-dependent Hamiltonian $H(t)$~\cite{tNQS_Bohrdt,tNQS_Carleo}. However, existing approaches to modeling time-evolved wavefunctions share a common limitation: each computation or optimization is performed for a single instance of a driving protocol. When $H(t)$ is changed, the entire procedure must be repeated from scratch. In practice, however, one is often interested in time evolution under a \emph{family} of driving protocols: for example, when optimizing pulse sequences for state preparation or benchmarking quantum simulators.

A natural question, then, is whether we can move beyond this pointwise paradigm. Recent works~\cite{Transformer,rende2025foundation,zaklama2026large} on ``foundational neural quantum states'' have taken a step in this direction by demonstrating the ability of a unified model to represent ground states across a family of Hamiltonians. However, this transferability across static system parameters remains limited.
Generalization in this setting amounts to interpolation within a parameter space $\boldsymbol{\lambda} \in \mathbb{R}^d$, and the model needs only learn a function of finitely many parameters. 

In contrast, driving protocols in time-dependent systems are qualitatively different: they are \emph{functions} of time, and therefore elements of an infinite-dimensional space. Generalizing across such a space is no longer interpolation, but requires learning an \emph{operator} that maps between function spaces. This is the setting of operator learning~\cite{NeuralOperator_general, FNO_Li, NeuralOperator2}, and it requires a fundamentally novel architecture and conceptual framework, which, to the best of our knowledge, have yet to be developed.

In this work, we bridge this gap by developing a foundational model that transfers across functional spaces of driving protocols. Our model takes a driving protocol $H(t)$ as input and returns the corresponding time-evolved quantum state $\ket{\psi(t)}$ as output. Rather than solving for the time evolution separately for each trajectory, the model, once trained, can predict dynamics under any protocol through a single forward pass. Our approach therefore represents a paradigm shift from \emph{solving} the Schr\"odinger equation to \emph{learning how to solve} it.

Concretely, we introduce the \emph{Neural Operator Quantum State (NOQS)}, a hybrid architecture that combines a transformer-based autoregressive wavefunction ansatz with a neural operator that maps between functional spaces. These two components process, respectively, the discrete many-body Hilbert space and the continuous temporal structure of the driving protocol. The NOQS model is trained in a self-supervised manner, without requiring external data.

We validate our approach on the two-dimensional transverse-field Ising model (TFIM) with time-dependent longitudinal and transverse fields. Trained on an ensemble of random driving protocols, the model generalizes accurately not only to in-distribution instances of the driving protocol, but also to qualitatively different out-of-distribution functional forms, such as Gaussian pulses and ramps. Furthermore, we show that the NOQS model can be trained on a coarse temporal grid and make accurately inference on a denser discretization, at times never encountered during training. The NOQS can be further efficiently fine-tuned in a protocol-specific manner, leading to improved accuracy across observables at all times, while requiring only a small number of local measurements.

Our work introduces the concept of operator learning over the space of driving protocols for quantum many-body dynamics. As evidenced by its ability to generalize to out-of-distribution functional forms and transfer between discretizations, our model learns a genuine functional mapping rather than merely memorizing individual trajectories. This capability opens a two-way interface between computation and experiment. In one direction, a pre-trained NOQS can provide predictions for local observables under previously unseen driving protocols without requiring costly numerical simulations. In the other direction, NOQS can incorporate sparse measurement data from experimental platforms and refine its prediction for the full quantum state at all times. Our framework therefore provides a practical bridge between numerical computation and experiment for driven quantum systems.

The rest of this paper is organized as follows. In Sec.~\ref{sec:NQS}, we review the basics of NQS. In Sec.~\ref{sec:noqs}, we introduce the NOQS model and describe the training procedure. In Sec.~\ref{sec:results}, we demonstrate the accuracy and transferability of the NOQS framework through numerical results on the two-dimensional driven TFIM. Finally, in Sec.~\ref{sec:conclusion}, we summarize our findings and discuss future research directions.

%alternative for a wide range of physical systems and problems~\cite{Donatella_autoregressive_dynamics,luodi_autoregressive_open, sharir2020_autoregressive,tNQS_Bohrdt,tNQS_Carleo, nqs_review1, nqs_review2}.
%The central idea of NQS is to approximate quantum many-body wavefunctions using a neural network as a variational ansatz~\cite{carleo2017}. 

\section{Neural-Network Quantum States \label{sec:NQS}}
In this section, we begin by reviewing the Neural-Network Quantum States (NQS), which form the foundation of our approach. We introduce the neural-network based variational ansatz, the autoregressive sampling scheme, and the variational Monte Carlo (VMC) framework used to evaluate expectation values during training and inference.

For concreteness, we will focus on quantum systems consisting of spin-half ($S=1/2$) degrees of freedom, whose canonical basis consists of product states of $S_z$, namely $\ket{\bm{\sigma}}\equiv\ket{\sigma_1,\sigma_2,\dots}$, with $\sigma_i\in\{-1,+1\}$ indicating the eigenvalues of $2S_z$.  For any spin configuration $\bm{\sigma}$, the wavefunction $\psi(\bm{\sigma})\equiv \langle\bm{\sigma}|\psi\rangle$ of a quantum state $\ket{\psi}$ takes a complex number. In other words, the wavefunction can be viewed as a function that maps from the space of configurations to $\mathbb{C}$:
\begin{equation}
    \psi : \{-1, +1 \}^N \rightarrow \mathbb{C},
\end{equation}
and each amplitude can be decomposed into two real numbers, an amplitude and a phase:
\begin{equation}
    \psi(\bm{\sigma}) = \sqrt{p(\bm{\sigma})} e^{i\phi(\bm{\sigma})},
\end{equation}
where both $p(\bm{\sigma})$ and $\phi(\bm{\sigma})$ are outputted by the neural networks. For stability of training and learning, in practice one often parameterizes the natural logarithm of the wavefunction,
\begin{equation}
    \log(\psi(\bm{\sigma})) = \frac{1}{2} \log( p(\bm{\sigma})) + i\phi(\bm{\sigma}).
\end{equation}

Early NQS architectures were based on restricted Boltzmann machines (RBMs) due to their analytical tractability~\cite{carleo2017, RBM_TN,RBM_review,RBM_symmetry, RBM_symmetry2}. More recently, autoregressive architectures such as recurrent neural networks (RNNs) and transformers have attracted considerable interest, owing to their ability to perform exact sampling~\cite{luodi_autoregressive_open, sharir2020_autoregressive, Donatella_autoregressive_dynamics, autoregressive-hubbard, barrett2022autoregressive, autoregressive-quantumnumber}. A key structural property exploited by autoregressive NQS is the factorization of the Born probability distribution. For any spin configuration $\bm{\sigma} = (\sigma_1, \sigma_2, \dots, \sigma_N)$, the joint probability of this configuration can be written as a product of conditional probabilities:
\begin{equation}
 p(\bm{\sigma}) = \prod_{i=1}^N  p(\sigma_i|\sigma_1, \dots,\sigma_{i-1})
 \label{eq:factorization}
\end{equation}
where $p(\cdot | \cdot)$ denotes the conditional probability of the $i$th spin, given all preceding spins. The phase is factorized in an analogous way. The conditional probabilities are produced by the autoregressive network.

This factorization has several important consequences. First, it enables exact, unbiased sampling: spins are generated sequentially, or \textit{autoregressively}, where each individual $\sigma_i$ is drawn from its own conditional distribution, yielding independent and identically distributed samples from $p(\bm{\sigma})$, without requiring Markov chain Monte Carlo. Second, it avoids direct normalization over the exponentially large configuration space; instead, the conditional probabilities are individually normalized by construction. These properties stabilize training, particularly for larger systems, and make autoregressive models a suitable ansatz to variationally represent quantum states.

Given samples autoregressively drawn from the Born distribution $p(\bm{\sigma}) = |\psi(\bm{\sigma})|^2$, expectation values, which are intractable to compute exactly for larger systems, can now be estimated stochastically. The expectation value of a general operator $O$ to the quantum state $\ket{\psi}$ can be estimated as:
\begin{equation}
    \langle O \rangle
    =
    \sum_{\bm{\sigma}}
    p(\bm{\sigma})\;
    O_{\mathrm{loc}}(\bm{\sigma}),
\end{equation}
where
\begin{equation}
    O_{\mathrm{loc}}(\bm{\sigma})
    =
    \frac{
        \bra{\bm{\sigma}} O \ket{\psi}
    }{
        \braket{\bm{\sigma}}{\psi}
    }
\end{equation}
is the local estimator. In practice, the sum is replaced by a sample average over configurations.

The variational ground state is obtained by optimizing the NQS parameters to minimize the sample average of the local energy, which is defined as
\begin{equation}
  E_{\mathrm{loc}}(\bm{\sigma}) = \sum_{\bm{\sigma}'} H_{\bm{\sigma}\bm{\sigma}'} \frac{\psi(\bm{\sigma}')}{\psi(\bm{\sigma})}.
  \label{eq:energy_estimator}
\end{equation}
This can be evaluated efficiently for local Hamiltonians, since only a polynomial number of configurations $\bm{\sigma}'$ yield non-trivial overlap with $\bm{\sigma}$ i.e. have non-vanishing matrix elements $H_{\bm{\sigma}\bm{\sigma}'}$.

\section{Neural Operator Quantum State \label{sec:noqs}}
\subsection{Problem Formulation}
\label{sec:problem}
The time evolution of a quantum many-body state is governed by the Schr\"odinger equation:
\begin{equation}
    i\partial_t |\psi(t)\rangle = H(t) |\psi(t)\rangle,
    \label{eq:schrodinger}
\end{equation}
where the Hamiltonian $H(t)$ is generally time-dependent. Formally, the time-evolved state can be written as a time-ordered exponential acting on the initial state $|\psi(0)\rangle$:
\begin{equation}
    |\psi(t)\rangle = \mathcal{T} \exp\!\left( -i \int_0^t dt'\, H(t') \right) |\psi(0)\rangle,
    \label{eq:time_evolution}
\end{equation}
where $\mathcal{T}$ denotes time ordering. Computing time-evolved states is a central challenge in quantum many-body physics, particularly for driven systems, where the Hamiltonian varies continuously in time. One must not only represent an exponentially large quantum state, but track its evolution under an arbitrary time-dependent protocol.

Previous approaches to this problem, such as exact diagonalization, tensor networks, Trotterized circuits, or time-dependent neural quantum states, share a common limitation. Namely, these approaches compute $\ket{\psi(t)}$ for a \emph{single trajectory} $H(t)$. If the driving protocol changes, the computation or optimization must be performed again. This point-wise paradigm presents a severe bottleneck in settings where one needs to track time evolution of quantum states across a family of time-dependent Hamiltonians: for instance, when scanning control parameters in a quantum simulator, or optimizing a driving protocol for state preparation.

In this work, we propose a fundamentally different approach. Rather than studying time evolution under individual trajectories, we learn the \emph{solution operator} itself. We introduce the Neural \textit{Operator} Quantum State (NOQS), which maps from driving protocols to time-evolved quantum states, given a \textit{fixed} initial state $\ket{\psi_0}$. More precisely, the NOQS is an operator $\mathcal{F}_{(\theta,\eta)}$ that depends on parameters $(\theta,\eta)$: 
\begin{equation}
    \mathcal{F}_{(\theta,\eta)} : H(t) \longmapsto \psi_\theta(\boldsymbol{\sigma}; \mathcal{N}_{\eta}[H(t)]),
    \label{eq:operator_mapping}
\end{equation}
where $\psi_\theta(\boldsymbol{\sigma} ;\mathcal{N}_{\eta}[H(t)])$ is the NOQS ansatz for the time-dependent wavefunction amplitude $\langle \boldsymbol{\sigma} | \psi(t)\rangle$, and we require that $\psi_\theta(\boldsymbol{\sigma};\mathcal{N}_\eta[H(t)])(t=0) \equiv \langle \boldsymbol{\sigma}|\psi_0\rangle$ at the starting time, for all $H(t)$. At a high level, $\psi_\theta$ is an NQS introduced in Sec.~\ref{sec:NQS}, conditioned on the parameters $\mathcal{N}_\eta[H(t)] \equiv M(t)$. The parameters $M(t)$ are called \textit{context tokens}: they are time-dependent vector functions, obtained from mapping the driving protocols $H(t)$ via a neural operator $\mathcal{N}_\eta$ parametrized by $\eta$. The explicit definitions of this hybrid architecture will be introduced in more detail in Sec.~\ref{sec:architecture}.
%We assumed we have explicitly spelled out the dependence of $\ket{\psi(t)}$ on the driving protocol $H(t)$, 
We note that the mapping in Eq.~\ref{eq:operator_mapping} is between \emph{function spaces}: the input and output are both functions of time. Learning this operator is the central goal of our work.

The conceptual shift, then, is from \emph{solving} to \emph{learning to solve} the Schr\"odinger equation. Training the operator $\mathcal{F}_{\theta,\eta}$ allows for predicting time evolution for not just one instance of $H(t)$, but instead over the full space of driving protocols. The optimization of $\mathcal{F}_{\theta,\eta}$ is performed once, over a family of Hamiltonians, after which the time-evolution under any new, unseen protocol is obtained in a single forward pass through the network, with no further training needed. This amortization of pre-training computational effort is what distinguishes our approach from both conventional solvers and existing time-dependent NQS methods. In spirit, our approach is analogous to how foundation models in machine learning (such as Large Language Models) are capable of performing well, over a distribution of tasks.

In order to train such a model, we need an architecture that can naturally handle the two degrees of freedom in the problem: the \emph{continuous} temporal structure of the driving protocol $H(t)$, and the \emph{discrete} structure of the spin configurations. In the following subsection, we introduce the NOQS architecture, which achieves this separation through a hybrid structure; coupling a Fourier Neural Operator, which processes the driving protocol in the time-frequency domain, to a transformer-based quantum state ansatz that represents the many-body wavefunction.

\subsection{Model Architecture \label{sec:architecture}}
\begin{figure*}[t]
    \centering
    \resizebox{\linewidth}{!}{%
    %\documentclass[tikz,border=10pt]{standalone}
%\usetikzlibrary{positioning, shapes.geometric, arrows.meta, calc, fit, shadows, backgrounds}
%\usepackage{comment}

% --- Color Palette ---
\definecolor{boxblue}{RGB}{218, 232, 252}
\definecolor{boxgreen}{RGB}{213, 232, 212}
\definecolor{boxorange}{RGB}{255, 204, 153}
\definecolor{boxpurple}{RGB}{225, 213, 231}
\definecolor{addyellow}{RGB}{255, 255, 153}
\definecolor{arrowblue}{RGB}{22, 79, 134}
\definecolor{arrowred}{RGB}{178, 34, 34}
\definecolor{panelgray}{RGB}{245, 245, 245}

%\begin{document}

\begin{tikzpicture}[
    font=\sffamily\bfseries\large, 
    line width=1.5pt,              
    % Node Styles
    base/.style={
        draw, 
        rounded corners=2pt, 
        align=center, 
        minimum height=0.9cm, 
        minimum width=2.8cm, 
        inner sep=4pt
    },
    blockstack/.style={
        base, 
        fill=boxorange, 
        double copy shadow={shadow xshift=3pt, shadow yshift=3pt, fill=boxorange}
    },
    add/.style={
        draw, 
        circle, 
        fill=addyellow, 
        minimum size=0.8cm, 
        font=\huge\bfseries
    },
    tensor/.style={
        base, 
        fill=boxgreen, 
        double copy shadow={shadow xshift=3pt, shadow yshift=3pt, fill=boxgreen}
    },
    % Arrow Styles
    stdarrow/.style={-{Stealth[length=3mm]}, line width=1.5pt},
    thickblue/.style={-{Stealth[length=5mm]}, color=arrowblue, line width=4pt},
    thickred/.style={-{Stealth[length=4mm]}, color=arrowred, line width=2.5pt}
]

% =========================================================
% PANEL (a): TRANSFORMER BACKBONE (Now on Left)
% =========================================================

\begin{scope}[xshift=0cm, yshift=-0.5cm]

    % Inputs
    \node[base, fill=boxblue, minimum width=2.2cm, font=\bfseries\Large] (embed) at (-1.9, 1.5) {Embedding};
    \node[base, fill=boxgreen, minimum width=2.2cm, font=\bfseries\Large] (pos) at (2.3, 1.5) {Pos. Encoding};
    
    % Sigma Inputs feeding BOTH
    \node[below=0.9cm of embed, xshift=1.9cm, font=\bfseries\Huge] (sig1) {$\sigma_1, \sigma_2, \dots, \sigma_N$};
    \draw[stdarrow] (sig1.north) -- (embed.south);
    \draw[stdarrow] (sig1.north) -- (pos.south);

    % Add
    \node[add, above=0.6cm of embed, xshift=1.9cm] (add_b) {+};
    \draw[stdarrow] (embed.north) |- (add_b.west);
    \draw[stdarrow] (pos.north) |- (add_b.east);

    % Decoder Stack
    \node[blockstack, above=0.8cm of add_b, minimum height=2.5cm, minimum width=4.5cm,font=\bfseries\Large] (decoder) {Decoder Layers};
    \draw[stdarrow] (add_b) -- (decoder);

    % Unembedding
    \node[base, fill=boxpurple, above=0.8cm of decoder,font=\bfseries\Large] (unembed) {Unembedding};
    \draw[stdarrow] (decoder) -- (unembed);
    
    % --- Split Output ---
    % Softmax -> p
    \node[base, fill=boxorange, above left=0.8cm and -1.0cm of unembed, minimum width=1.8cm, font = \bfseries\Large] (softmax) {Softmax};
    \node[above=0.4cm of softmax, font=\bfseries\Huge] (prob) {$\log(p({\bm{\sigma}}; t))$};
    \draw[stdarrow] (unembed.north) -- ++(0,0.3) -| (softmax.south);
    \draw[stdarrow] (softmax) -- (prob);

    % Phase phi
    \node[above right=2.2cm and -0.8cm of unembed, font=\bfseries \Huge] (phase) {$\phi(\bm{\sigma}; t)$};
    \draw[stdarrow] (unembed.north) -- ++(0,0.3) -| (phase.south);

\node[above left=-0.1cm and -0.6cm of prob.north west, font=\huge\bfseries] {(a)};

\end{scope}

% =========================================================
% PANEL (b): DETAILED LAYER (Center - "The Zoom In")
% =========================================================

\begin{scope}[xshift=9.0cm, yshift=-1.0cm]

    % Background Box for the Detail
    \node[draw, fill=panelgray, rounded corners=10pt, minimum width=7cm, minimum height=11.5  cm, anchor=south] (bg_detail) at (0,0) {};

    % Input Arrow
    \coordinate (c_in) at (0, -0.5);

    % 1. Self Attention
    \node[base, fill=boxorange, above=1.2cm of c_in, minimum width=5cm, font=\bfseries\Large] (self_attn) {Masked Self-Attn};
    \node[add, above=0.8cm of self_attn] (add1) {+};
    \node[right=0.1cm of add1, font=\Large, align=left,font=\bfseries\large] {Add \&\\Norm};

    % 2. Cross Attention (Receives Context Directly)
    \node[base, fill=boxorange, above=0.8cm of add1, minimum width=5cm,font=\bfseries\Large] (cross_attn) {Cross-Attention};
    \node[add, above=0.8cm of cross_attn] (add2) {+};
    \node[right=0.1cm of add2, font=\Large, align=left,font=\bfseries\large] {Add \&\\Norm};

    % 3. FFN
    \node[base, fill=boxpurple, above=0.8cm of add2, minimum width=5cm,font=\bfseries\Large] (ffn) {Feed-Forward};
    \node[add, above=0.8cm of ffn] (add3) {+};
    \node[right=0.1cm of add3, font=\Large, align=left,font=\bfseries\large] {Add \&\\Norm};

    % Output
    \coordinate (c_out) at (0, 12.8);

    % --- Main Trunk (Thick Blue) ---
    \draw[thickblue] (c_in) -- (self_attn); 
    \draw[thickblue] (self_attn) -- (add1);
    \draw[thickblue] (add1) -- (cross_attn);
    \draw[thickblue] (cross_attn) -- (add2);
    \draw[thickblue] (add2) -- (ffn);
    \draw[thickblue] (ffn) -- (add3);
    \draw[thickblue] (add3) -- (c_out);

    % --- Residuals (Red) ---
    \draw[thickred, rounded corners=5pt] ($(c_in)!0.7!(self_attn.south)$) -- ++(-3.0,0) |- (add1.west);
    \draw[thickred, rounded corners=5pt] ($(add1.north)!0.5!(cross_attn.south)$) -- ++(-3.0,0) |- (add2.west);
    \draw[thickred, rounded corners=5pt] ($(add2.north)!0.5!(ffn.south)$) -- ++(-3.0,0) |- (add3.west);

\node[above right=1.0cm and 0.0cm of bg_detail.north west, font=\huge\bfseries] {(b)};
\end{scope}

% =========================================================
% PANEL (c): PROTOCOL ENCODER / FNO (Now on Right)
% =========================================================

\begin{scope}[xshift=17.5cm, yshift=-0.0cm]

    % --- Waveform Input ---
    \node (waveform) at (0, 0) {};
    \draw[arrowblue, line width=2.5pt, smooth] plot coordinates {
        (-1.2, 0) (-0.8, 0.4) (-0.4, -0.3) (0, 0.5) (0.4, -0.2) (0.8, 0.3) (1.2, 0)
    };
    \node[below=0.3cm of waveform,font=\bfseries\Huge] {$H(t)$};

    % --- Lifting Layer ---
    \node[base, fill=boxblue, above=1.0cm of waveform,font=\bfseries\Large] (lift) {Lifting};
    \draw[stdarrow] (0, 0.3) -- (lift);

    % --- FNO Layers ---
    \node[blockstack, above=0.8cm of lift, minimum height=2.5cm, minimum width=4.5cm,font=\bfseries\Large] (fno) {FNO Layers};
    \draw[stdarrow] (lift) -- (fno);

    % --- Projection & Context ---
    \node[base, fill=boxblue, above=0.9cm of fno,font=\bfseries\Large] (proj) {Projection};
    \draw[stdarrow] (fno) -- (proj);

    \node[tensor, above=0.7cm of proj, minimum height=1.0cm, font=\bfseries\Large] (rawcontext) {Raw Context\\$\widetilde{M}(t)$};
    \draw[stdarrow] (proj) -- (rawcontext);

    \node[tensor, above=0.7cm of rawcontext, minimum height=1.0cm, font=\bfseries\Large] (context) {Context\\$M(t)$};
    \draw[stdarrow] (rawcontext) -- (context);

\node[above left=2.1cm and 0.2cm of rawcontext.north west, font=\huge\bfseries] {(c)};
\end{scope}

% =========================================================
% GLOBAL CONNECTIONS
% =========================================================

% 1. CONTEXT BUS: (c) Right -> (b) Center
% The context M(t) flows from the Right (FNO) into the middle panel's Cross-Attention
\draw[thickred, line width=3pt, rounded corners=15pt, ->] 
    (context.west) -- ++(-2.0, 0) 
    coordinate (mid_point) 
    |- (cross_attn.east) 
    node[midway, above, xshift=-0.7cm, yshift=-1.0cm, font=\bfseries\Large, text=arrowred] {Condition};

% 2. ZOOM RELATIONSHIP: (b) Center -> (a) Left
% Draw dashed lines from the single layer in the middle to the stack on the Left (Transformer)
\draw[dashed, color=black!60, line width=1.5pt] 
    (bg_detail.north west) -- (decoder.north east);
\draw[dashed, color=black!60, line width=1.5pt] 
    (bg_detail.south west) -- (decoder.south east);

\end{tikzpicture}

%\end{document}
}%
    \caption{\textbf{(a)} Illustration of the architecture for transformer-based ansatz for quantum state. The input is a spin configuration, $\bm{\sigma} = \{ \sigma_1, \sigma_2, \dots, \sigma_N\}$, $\sigma_i \in \{-1, +1\}$. The physical spins are first mapped to a $d_e$-dimensional latent space via an embedding operation, augmented by positional encodings that contain information about their spatial location. The latent space representations pass through $L_T$ decoder layers. The final latent state is projected by the unembedding layer to yield the log-amplitude $\log(p(\bm{\sigma}; t))$ and phase $\phi(\bm{\sigma}; t)$.
    \textbf{(b)} Internal structure of each decoder block, showing the masked multi-head self-attention mechanism, residual connections, feed-forward network, and importantly the cross-attention mechanism that attends to temporal information. \textbf{(c)} Time-dependent driving protocols $H(t)$ are processed by the Fourier Neural Operator (FNO) and projected to raw context tokens $\widetilde{M}(t)$, which are then offset by their initial values (Eq.~\ref{eq:rawtocontextvec}) to respect the initial condition. The transformer wavefunction ansatz attends to the processed context tokens $M(t)$ through a cross-attention mechanism.}
    \label{fig:transformer}
\end{figure*}
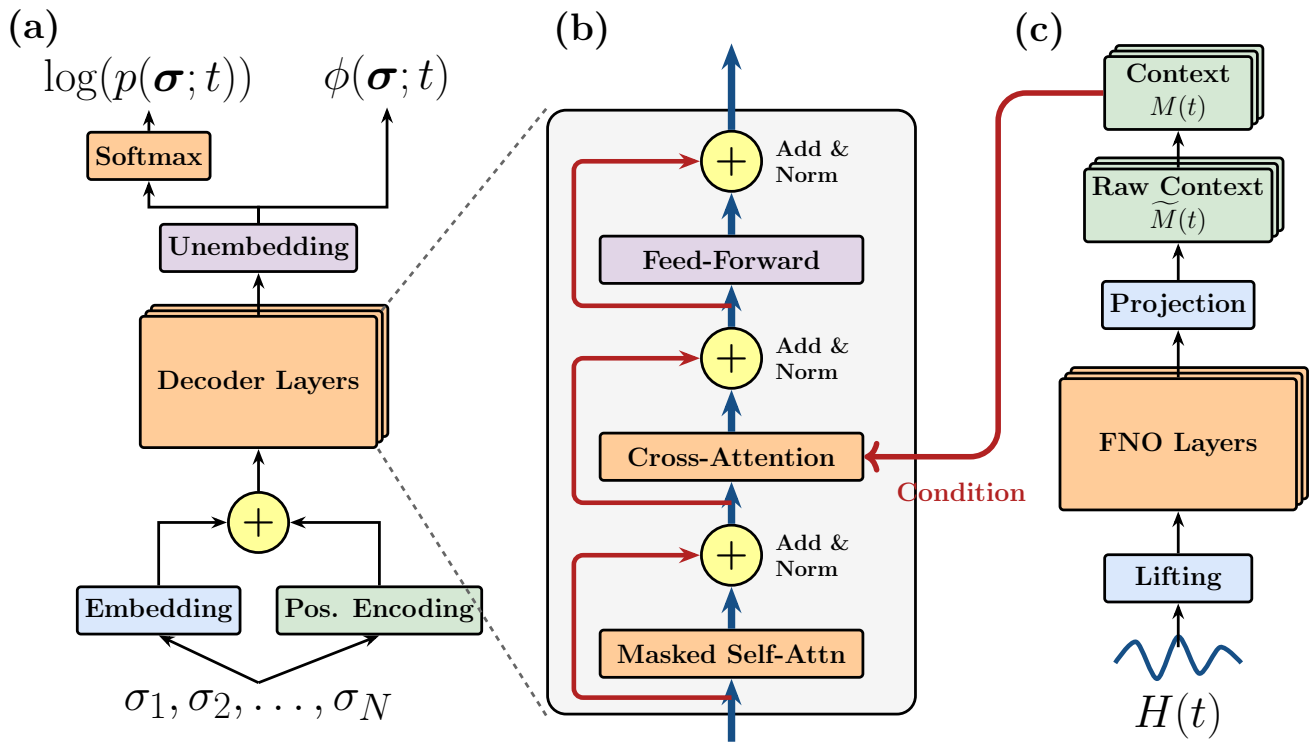

The backbone of our NOQS model is a transformer-based, autoregressive representation of the quantum many-body wavefunction. The transformer architecture~\cite{vaswani2017attention}, with its all-to-all self-attention mechanism, has been shown to be a powerful ansatz for representing a wide range of spin and fermionic quantum states~\cite{barrett2022autoregressive,Transformer,autoregressive-hubbard, transformer_wf_2, transformer_wf_3}. We now begin with a description concerning the architectural details of the proposed transformer model, as illustrated in Fig.~\ref{fig:transformer}(a).

The inputs to the transformer are spin configurations $\bm{\sigma} = (\sigma_1, \sigma_2, \dots, \sigma_N)$. Each physical spin $\sigma_i \in \{+1, -1\}$ is first mapped to a token in a $d_e$-dimensional latent space via an embedding layer:
\begin{equation}
    \mathbf{e}_i = W_E\, \sigma_i, \quad W_E \in \mathbb{R}^{d_e}
    \label{eq:embedding}
\end{equation}
where $W_E$ is a learnable weight matrix shared across all sites $i$. This produces an embedding matrix $E \in \mathbb{R}^{N \times d_e}$ for any configuration of $N$ spins.

Since the embedding layer is identical across all spins, we also need to inject information about the spatial locations of the spins. Therefore, we augment each token with a (learnable) positional encoding $\mathbf{p}_i \in \mathbb{R}^{d_e}$ that carries information about the geometry of the underlying lattice. The input to the decoder layers is therefore:
\begin{equation}
    \mathbf{x}_i = \mathbf{e}_i + \mathbf{p}_i, \qquad i = 1, \ldots, N,
    \label{eq:pos_encoding}
\end{equation}
yielding an input matrix $X \in \mathbb{R}^{N \times d_e}$ consisting of $\mathbf{x}_i$ as row vectors.

The representations $\{\mathbf{x}_i\}$ are sequentially processed by a stack of $L_T$ decoder blocks. Each decoder block consists of three sub-layers with residual connections and layer normalization, as shown in Fig.~\ref{fig:transformer}(b). Next we discuss this structure of the decoder blocks; for simplicity of notation, we suppress the index of the decoders.

In each decoder block, the first sub-layer that $X$ passes through is a \emph{masked, multi-head self-attention} mechanism. For each of the $n_h$ attention heads, the input $X \in \mathbb{R}^{N \times d_e}$ is projected to queries, keys, and values:
\begin{equation}
    Q^{(s)} = X\, W_Q^{(s)}, \quad K^{(s)} = X\, W_K^{(s)}, \quad V^{(s)} = X\, W_V^{(s)},
    \label{eq:self_attn_proj}
\end{equation}
where $W_Q^{(s)}, W_K^{(s)}, W_V^{(s)} \in \mathbb{R}^{d_e \times d_h}$ and $d_h = d_e / n_h$ is the embedding dimension per attention head. Here the superscript $(s)$ is an index for the attention heads. The self-attention output for each head is:
\begin{equation}
    \mathrm{Attn}^{(s)} = \mathrm{softmax}\!\left( \frac{Q^{(s)} {K^{(s)}}^T}{\sqrt{d_h}} + \mathcal{M} \right) V^{(s)},
    \label{eq:self_attn}
\end{equation}
where $\mathrm{softmax}$ is applied row-wise. For any vector $A \in \mathbb{R}^{d}$, the softmax is defined as:
\begin{equation}
    \mathrm{softmax}(A)_{i} = \frac{\exp(A_{i})}{\sum_{k=1}^{d} \exp(A_{k})},
    \label{eq:softmax}
\end{equation}
where $A_i$ denotes the $i$th entry of the vector. The softmax, similar to the partition function, assigns a (normalized) weight to each entry of the vector. Inside the softmax argument, we impose a causal mask $\mathcal{M} \in \mathbb{R}^{N \times N}$ defined as:
\begin{equation}
    \mathcal{M}_{ij} = 
    \begin{cases}
        0 & \text{if } j \leq i, \\
        -\infty & \text{if } j > i.
    \end{cases}
    \label{eq:causal_mask}
\end{equation}

This mask enforces causality and the autoregressive property, and is crucial for modeling physical systems~\cite{autoregressive-hubbard, luodi_autoregressive_open, krylov_transformer, sharir2020_autoregressive}. It ensures that the representation of spin $\sigma_i$ attends \textit{only} to spins $\sigma_1, \ldots, \sigma_{i-1}$ that precede it, thereby preserving the conditional factorization of Eq.~\eqref{eq:factorization}. The outputs of all heads are concatenated,
\begin{equation}
    \mathrm{SelfAttn}(X) = \mathrm{Concat}\!\left(\mathrm{Attn}^{(1)}, \ldots, \mathrm{Attn}^{(n_h)}\right) W_O + b_O,
    \label{eq:self_attn_concat}
\end{equation}
and projected by $W_O \in \mathbb{R}^{d_e \times d_e}$ along with a bias vector $b_O \in \mathbb{R}^{d_e}$, followed by a residual connection and layer normalization.

Following the self-attention block is the \emph{cross-attention} mechanism that incorporates temporal information, processed by the Fourier Neural Operator (FNO). We will discuss how the FNO processes temporal information, as well as details regarding the cross attention mechanism, later in this section. 

The third and final component in each decoder block is a position-wise feed-forward network (FFN), applied identically to each token:
\begin{equation}
    \mathrm{FFN}(\mathbf{x}) = W_2\, \sigma(W_1\, \mathbf{x} + \mathbf{b}_1) + \mathbf{b}_2,
    \label{eq:ffn}
\end{equation}
where $W_1 \in \mathbb{R}^{d_f \times d_e}$, $W_2 \in \mathbb{R}^{d_e \times d_f}$, $\sigma(\cdot)$ is a non-linear activation function (namely GeLU), and $d_f$ is the feed-forward hidden dimension. This is also followed by a residual connection and layer normalization.

Next, we discuss how our architecture handles temporal information. We use neural operators to process the time-dependent driving protocol $H(t)$ and distill it into a set of \textit{context tokens} that condition the transformer wavefunction. In contrast to traditional neural networks, which map between finite-dimensional vectors, neural operators are generalizations that describe mappings between function spaces defined over a domain of interest~\cite{NeuralOperator_general}. Here, the relevant domain is the time interval $\mathcal{D} = (0, T_{\mathrm{max}})$. The neural operator in our architecture is therefore an operator mapping from $H(t)$ to $M(t)$:
\begin{equation}
    \mathcal{N}: \, H(t) \longrightarrow {M}(t).
\end{equation}
%\Yang{Should be $M(t)$, not the raw M. I want to denote $\mathcal{N}$ as the overall neural operator from H to the final M.}
%where $\widetilde{M}(t)$ is the set of (raw) time-dependent contexte tokens.

Specifically, we choose the Fourier Neural Operator (FNO) in our architecture due to its natural advantages in this setting: the FNO parameterizes the integral kernel directly in the frequency domain and has been shown to be discretization invariant~\cite{FNO_Li, FNOFloquet} (i.e., its performance depends minimally on the temporal resolution.) Furthermore, FNOs are parameter-efficient and computationally fast due to implementation using (inverse) Fast Fourier Transform (FFT)~\cite{FNO_Li, NeuralOperator_general}. Crucially, working in the frequency domain also provides an elegant route to computing time derivatives, as we will discuss in Sec.~\ref{sec:training}.

At each discretized time point $t_j$ ($j = 1, \ldots, N_t$), the input to the FNO is a vector of time-dependent coefficients of the Hamiltonian, expanded in the Pauli string basis: $\mathbf{H}(t_j) \in \mathbb{R}^{d_{\mathrm{in}}}$, where $d_{\mathrm{in}}$ is the number of time-dependent terms (equivalently, number of driving fields). A pointwise lifting layer first projects these coefficients to a higher-dimensional feature space of dimension $d_v$:
\begin{equation}
    \mathbf{v}^{(0)}(t_j) = W_L\, \mathbf{H}(t_j) + \mathbf{b}_L, \quad W_L \in \mathbb{R}^{d_v \times d_{\mathrm{in}}},\quad \mathbf{b}_L \in \mathbb{R}^{d_v},
    \label{eq:lifting}
\end{equation}
producing a lifted representation $V^{(0)} \in \mathbb{R}^{N_t \times d_v}$.

The core of the FNO is a stack of $L_F$ Fourier layers. In each layer $\ell$, the representation $V^{(\ell)} \in \mathbb{R}^{N_t \times d_v}$ is first transformed to the frequency domain via a Fast Fourier Transform (FFT) along the time axis. The transform is truncated to retain only the dominant $k_{\mathrm{max}}$ Fourier modes, yielding $\hat{V}^{(\ell)} \in \mathbb{R}^{k_{\mathrm{max}} \times d_v}$. A learnable weight matrix then mixes these frequency modes:
\begin{equation}
    \hat{V}'^{(\ell)} = R^{(\ell)}\, \hat{V}^{(\ell)}, \qquad R^{(\ell)} \in \mathbb{R}^{d_v \times d_v},
    \label{eq:fourier_mixing}
\end{equation}
where $R^{(\ell)}$ is applied identically to each of the $k_{\mathrm{max}}$ modes. 

Since multiplication in the frequency domain corresponds to convolution in the time domain, each Fourier layer learns nonlocal temporal correlations and couples information of $H(t)$ from different times. The filtered modes are transformed back to the time domain via an inverse FFT. A pointwise linear bias $W_s^{(\ell)} \in \mathbb{R}^{d_v \times d_v}$ is added, followed by a nonlinear activation $\sigma$:
\begin{equation}
    V^{(\ell+1)} = \sigma \left(\mathcal{F}^{-1}\!\left(\hat{V}'^{(\ell)}\right) + V^{(\ell)}\, W_s^{(\ell)} \right).
    \label{eq:fourier_layer}
\end{equation}

After $L_F$ Fourier layers, the output is projected to a set of $N_c$ context tokens. Concretely, at each point of time $t$, we lift the $d_v$-dimensional output to be $N_c \times d_e$ dimensional through a two-layer, feed-forward network. Finally, the output of this projection operation is reshaped to yield $N_c$ raw context vectors, each of dimension $d_e$. This reshaping ensures that the dimension of each token matches the embedding dimension of the transformer backbone, specifically:
\begin{equation}
    \widetilde{M}_1(t), \ldots, \widetilde{M}_{N_c}(t), \qquad \widetilde{M}_i(t) \in \mathbb{R}^{d_e} \quad \forall\, i,
    \label{eq:context_tokens}
\end{equation}

Importantly, to satisfy the physical constraint that time evolution starts from a fixed initial state for all driving protocols $H(t)$, we fix the initial context tokens, $M(t=0)$. We then offset the raw context tokens $\widetilde{M}(t)$ outputted from the FNO by the raw initial values $\widetilde{M}(0)$:
\begin{equation}
    M(t) \equiv \mathcal{N}[H(t)](t) = M(0) + \widetilde{M}(t) - \widetilde{M}(0),
    \label{eq:rawtocontextvec}
\end{equation}
which explicitly defines the neural operator $\mathcal{N}$. We note that these processed context tokens are \emph{functions of time}: at each time $t$, they provide a $d_e$-dimensional summary of the driving protocol in latent space.

The cross-attention mechanism bridges the two branches of the architecture: it allows the transformer, which processes the discrete spin degrees of freedom, to query the temporal information encoded by the FNO. This coupling is what enables the overall model to predict the wavefunction conditioned on any driving protocol $H(t)$ at any time $t \in (0, T_\text{max})$. In our architecture, the FNO plays a role analogous to encoders in encoder-decoder transformer structure~\cite{vaswani2017attention}.

Concretely, cross-attention appears as the second sub-layer in each decoder block (Fig.~\ref{fig:transformer}b). Let $X' \in \mathbb{R}^{N \times d_e}$ denote the output of the self-attention sub-layer for a given block, and let $M(t) \in \mathbb{R}^{N_c \times d_e}$ be the context matrix produced by the neural operator $\mathcal{N}$ at time $t$. The transformer representations serve as queries, while the context tokens provide the keys and values. For each of $n_h$ attention heads:
\begin{equation}
    Q^{(c)} = W_Q^{(c)} X', \quad K^{(c)} = W_K^{(c)} M(t), \quad V^{(c)} = W_V^{(c)} M(t),
    \label{eq:cross_attn_proj}
\end{equation}
where $W_Q^{(c)} \in \mathbb{R}^{d_e \times d_h}$, $W_K^{(c)} \in \mathbb{R}^{d_e \times d_h}$, and $W_V^{(c)} \in \mathbb{R}^{d_e \times d_h}$. Note that the super-script $(c)$ denotes the weight matrices in the \textit{cross}-attention mechanism. The output is:
\begin{equation}
    \mathrm{Attn}^{(c)} = \mathrm{softmax}\!\left( \frac{Q^{(c)} {K^{(c)}}^\top}{\sqrt{d_h}} \right) V^{(c)},
    \label{eq:cross_attn}
\end{equation}
where the attention matrix has shape $\mathbb{R}^{N \times N_c}$: each of the $N$ spin tokens attends to each of the $N_c$ context tokens. Unlike the self-attention mechanism, no causal mask is applied here, as every spin token should have access to the full temporal context. The heads are concatenated and projected, followed by a residual connection and layer normalization, as in the case of self-attention (Eq.~\ref{eq:self_attn_concat}).

We note that the cross-attention mechanism is not symmetric in $X$ and $M$: intuitively, the cross-attention allows each spin's latent representation to ``look up'' the relevant temporal context: how the driving fields are evolving at the current time $t$, and what temporal correlations the neural operator $\mathcal{N}_\eta$ has identified. Because the context tokens $M(t)$ vary continuously with $t$ (inheriting the smoothness of the FNO output), the transformer's predictions also vary smoothly in time, without requiring any explicit time discretization in the wavefunction ansatz.

In summary, our architecture couples a transformer-based variational wavefunction ansatz and a neural operator through cross attention. The two components handle the spin and physical degrees of freedom, respectively. The hyperparameters we used in the model are listed in Appendix.~\ref{app:hyperparameters}.

\subsection{Training Procedure}
\label{sec:training}

The trainable parameters of the NOQS consist of those in the transformer, the FNO, and the cross-attention projection matrices. Let $\{\theta, \eta \} \in \mathbb{R}^m$ denote the full set of parameters in the NOQS model, where $m$ is the total number of parameters. The objective of training is to find an optimal set of parameters $\{\theta^*, \eta^*\}$ that minimizes a loss function $\mathcal{L}$.

We define the loss function based on the Time-Dependent Variational Principle (TDVP)~\cite{TDVP_original, tdvp, TDVP_quantum}, which identifies the optimal trajectory within a variational manifold (defined by the parameters $\theta$), through minimizing the residual of the Schr\"odinger equation~\eqref{eq:schrodinger}:
\begin{equation}
    \mathcal{L}_{\mathrm{TDVP}} = \int dt\; \bigl\lVert
    \left( i\partial_t - H(t) \right) \, \psi_\theta(\mathcal{N}_{\eta}[H(t)]) \bigr\rVert^2.
    \label{eq:tdvp_loss}
\end{equation}
This loss reaches its minimum, $\mathcal{L}_{\mathrm{TDVP}} = 0$, if and only if $\psi_\theta(\mathcal{N}_{\eta}[H(t)])$ satisfies the Schr\"odinger equation exactly, under time evolution governed by $H(t)$.

For larger systems, it is generally infeasible to exactly evaluate $\mathcal{L}_\text{TDVP}$; instead, the loss must be evaluated stochastically. We rewrite the integrand using local estimators introduced in Sec.~\ref{sec:NQS}. Expanding the norm and inserting a resolution of identity, we can express the TDVP loss at a fixed time $t$ as:
\begin{align}
    \mathcal{L}_{\mathrm{TDVP}}(t) 
    &= \bra{\psi_\theta} (i\partial_t - H(t))^\dagger (i\partial_t - H(t)) \ket{\psi_\theta} \\
    &= \sum_{\bm{\sigma}} p_{\theta, \eta}(\bm{\sigma})\, \left| L_{\mathrm{loc}}(\bm{\sigma}, t) \right|^2,
    \label{eq:tdvp_local}
\end{align}
where the local estimator for Schr\"odinger residual is:
\begin{equation}
    L_{\mathrm{loc}}(\bm{\sigma}, t) 
    = i\,\partial_t \log \psi_\theta(\bm{\sigma};\mathcal{N}_{\eta}[H(t)]) - E_{\mathrm{loc}}(\bm{\sigma}, t).
    \label{eq:schrodinger_residual}
\end{equation}
Here $E_{\mathrm{loc}}(\bm{\sigma}, t)$ is the local energy estimator defined in Eq.~\eqref{eq:energy_estimator}, and $\partial_t \log \psi_\theta(\bm{\sigma};\mathcal{N}_{\eta}[H(t)])$ is the time derivative of the logarithmic wavefunction. Both terms are estimated at each step of training. In practice, we exploit the gauge freedom of a global phase in the physical state and minimize the variance of $|L_{\text{loc}}(\sigma,t)|^2$ at each $t$; for more details, see Appendix.~\ref{app:hyperparameters}.
%\Yang{Shall we say (in Appendix) that we practically minimize the variance of $|L_{loc}(\sigma,t)|^2$ at each $t$? We need to say if $L_{loc}$ at each $t$ is $\sigma$ independent, then it is enough to give the correct physical state, because physical state are defined up to a global phase.}

A key advantage of processing the driving protocol using the FNO is that the time derivative in Eq.~\ref{eq:schrodinger_residual} can be computed in frequency domain. Recall that the context tokens $M(t) \equiv \mathcal{N}_{\eta}[H(t)]$ are produced by the FNO. Consequently, they inherit a Fourier representation from the internal layers. For any such function $f(t) = \sum_k \hat{f}_k\, e^{i\omega_k t}$, its time derivative is:
\begin{equation}
    \partial_t f(t) = \sum_k (i\omega_k)\, \hat{f}_k\, e^{i\omega_k t},
    \label{eq:spectral_diff}
\end{equation}
which amounts to multiplying each Fourier coefficient by $i\omega_k$, a pointwise operation in frequency space. Since the wavefunction $\psi_\theta(\bm{\sigma}; M(t))$ depends on time \textit{entirely} through the context tokens $M(t)$ via the cross-attention mechanism, the chain rule yields an exact expression for $\partial_t \log \psi_\theta(\bm{\sigma};\, t)$ in terms of the Fourier coefficients, combined with automatic differentiation of the transformer ansatz. This avoids the numerical instabilities and dependence on discrete time grids with finite-difference approximations. Performing the derivative in frequency domain allows the NOQS to inherit the discretization invariance of the FNO and transfer across temporal discretizations, as we will show in Sec.~\ref{sec:discretization_invariance}.

As with any initial-condition PDE problem, it is crucial to ensure that the initial condition is enforced; i.e., to ensure that the output state at $t=0$ matches $\ket{\psi_0}$ exactly. We approach this in a few complementary ways. First, we pre-train the NOQS to condition it on the given $\ket{\psi_0}$. During training, we also supplement the TDVP loss with an anchor term that enforces the initial condition:
\begin{equation}
    \mathcal{L}_{\mathrm{anchor}} = \bigl\lVert \psi_\theta(\mathcal{N}_{\eta}[H(t)])(t=0) - \ket{\psi_0} \bigr\lVert^2,
    \label{eq:anchor_loss}
\end{equation}
which penalizes deviations of the learned state from the known initial state at $t = 0$. Training with only the TDVP term (Eq.~\ref{eq:schrodinger_residual}) might allow the initial state to drift during training, since the Schr\"odinger equation constrains only the time derivative. The anchor loss, therefore, is particularly helpful for stabilizing the global phase of the wavefunction, which the TDVP loss leaves undetermined modulo $2\pi$. The total loss function used during training is:
\begin{equation}
    \mathcal{L} = \mathcal{L}_{\mathrm{TDVP}} + \lambda_w\, \mathcal{L}_{\mathrm{anchor}},
    \label{eq:total_loss}
\end{equation}
where $\lambda_w$ is a hyperparameter controlling the strength of the initial-condition constraint. Along with the offset for the context token (Eq.~\ref{eq:rawtocontextvec}), which enforces the initial condition on the level of the neural operator, the anchor loss additionally ensures that the initial condition is respected at the wavefunction level.

%\Yang{Since in previous paragraphs, I already added M(0) is fixed for all H(t), should we remove this paragraph? Or just say one sentence on fixing M(0) in the previous paragraph regarding how to pretrain with the anchor loss. }Finally, we ensure that the context token $M(t)$ is identical at $t=0$, for \textit{any} driving protocol $H(t)$. Indeed, since the time-dependence of the NOQS comes entirely from the context tokens, ensuring $M(t=0)$ is fixed and unchanged during training is crucial for guaranteeing that the initial condition is respected.

At each training step, we sample a batch of $B$ Hamiltonian trajectories $\{H^{(b)}(t)\}_{b=1}^B$ from the training distribution. For each trajectory, we randomly select $K$ time points $\{t_k\}$, where $t_k \in (0, T_\text{max})$. Finally, at each time point, we draw $M$ spin configurations $\{\bm{\sigma}^{(m)}\}$ from the Born distribution $p_{\theta, \eta}(\bm{\sigma})$ via autoregressive sampling. The loss and its gradients are then estimated by averaging over all three stochastically:
\begin{equation}
    \mathcal{L} \approx \frac{1}{BKM} \sum_{b=1}^{B} \sum_{k=1}^{K} \sum_{m=1}^{M}
    \left| L_{\mathrm{loc}}\!\left(\bm{\sigma}^{(m)},\, t_k;\, H^{(b)}\right) \right|^2
    + \lambda_w\, \mathcal{L}_{\mathrm{anchor}}.
    \label{eq:mc_loss}
\end{equation}
Details about the parameters used in training are also included in Appendix.~\ref{app:hyperparameters}.

We emphasize that, unlike standard VMC or single-trajectory TDVP, our training loop samples over both spin configurations \emph{and} Hamiltonian trajectories at each step. This average over protocols, times, and configurations is what enables the model to learn the operator mapping of Eq.~\eqref{eq:operator_mapping} across the full function space, rather than optimizing for individual trajectories. We also highlight that the whole training process does not require any external data. The whole architecture is trained using a physically motivated loss function, namely deviation from the Schr\"odinger equation and the initial condition, in a self-supervised manner.

\section{Numerical results \label{sec:results}}

\subsection{TFIM with Transverse and Longitudinal Quench}
To validate our approach, we consider a system of spin-$1/2$ degrees of freedom on an $L_x \times L_y$ square lattice with open boundary conditions (OBC). Specifically, we focus on the two-dimensional transverse-field Ising model (TFIM) with time-dependent longitudinal and transverse fields, which has been numerically studied and can be engineered on various experimental platforms~\cite{TFIM, TFIM_quench, TFIM_experiment, TFIM_experiment2, TFIM_dynamics, TFIM_Rydberg}. The Hamiltonian reads:
\begin{equation}
    H(t) = J \sum_{\langle ij \rangle} Z_i Z_j + h_x(t) \sum_i X_i + h_z(t) \sum_i Z_i,
    \label{eq:hamiltonian}
\end{equation}
where $\langle ij \rangle$ denotes nearest-neighbor bonds on the square lattice, and $X_i$ ($Z_i$) is the Pauli-$x$ ($z$) operator on site $i$. The functions $h_x(t)$ and $h_z(t)$ are the time-dependent transverse and longitudinal fields, respectively. A nonvanishing longitudinal field $h_z(t)$ breaks the integrability of the model, making the dynamics nontrivial and precluding analytical solutions.

We fix $J = 1$ throughout. During training, the driving fields $h_x(t)$ and $h_z(t)$ are sampled from a family of smooth, random trajectories generated via truncated Fourier series:
\begin{equation}
    h_x(t) = h_{x0} + \sum_{m=1}^{n_{\max}} a_m \sin(m\omega\, t + \phi_m),
    \label{eq:driving_field}
\end{equation}
and similarly for $h_z(t)$. Here $n_{\max} = 10$ is the number of Fourier modes, $\omega = 10\,J$ sets the fundamental frequency, and $h_{x0} = 1.0\,J$ is a constant offset. The amplitudes $a_m$ are drawn from a normal distribution with standard deviation $0.6\,J / m^{3/2}$; the mode-dependent normalization suppresses higher harmonics and ensures smoothness, while the phases $\phi_m$ are sampled uniformly from $(0, 2\pi]$. The longitudinal field $h_z(t)$ is generated analogously but with a smaller amplitude scale ($0.05\,J$) and zero mean offset $h_{z0} = 0$. This parameterization produces a diverse yet smooth ensemble of driving protocols for both training and evaluation.

\begin{figure}[t]
    \centering
    \includegraphics[width=1\linewidth]{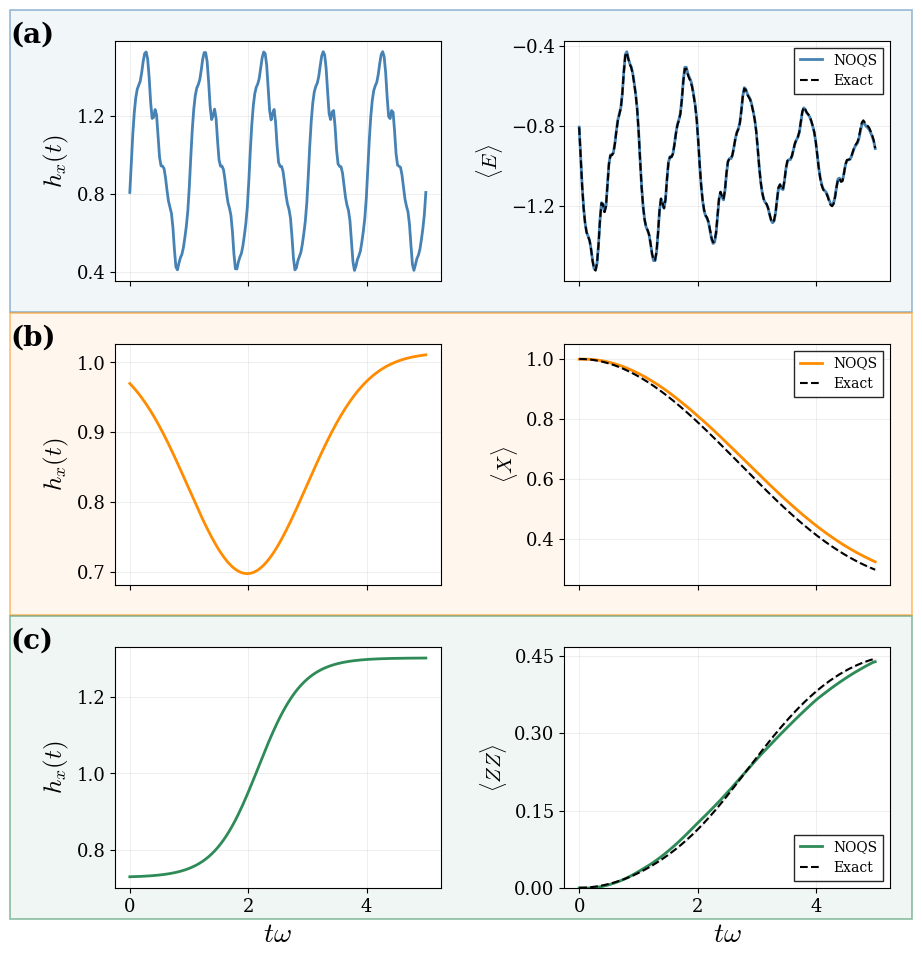}
    \caption{Transverse field $h_x(t)$ and expectation values of local observables for system size $4 \times 4$, benchmarked against exact numerical results. \textbf{(a)} Performance of NOQS on predicting $E(t)$ for in-distribution driving fields unseen during training. The NOQS also generalizes to out-of-distribution driving protocols; for \textbf{(b)} Gaussian pulses and \textbf{(c)} tanh ramps. The NOQS model predicts local observables accurately in all three cases, indicating the learning of time-evolved quantum state under a functional space of driving fields.}
    \label{fig:4by4}
\end{figure}

\begin{figure*}[t]
    \centering
    \includegraphics[width=0.75\linewidth]{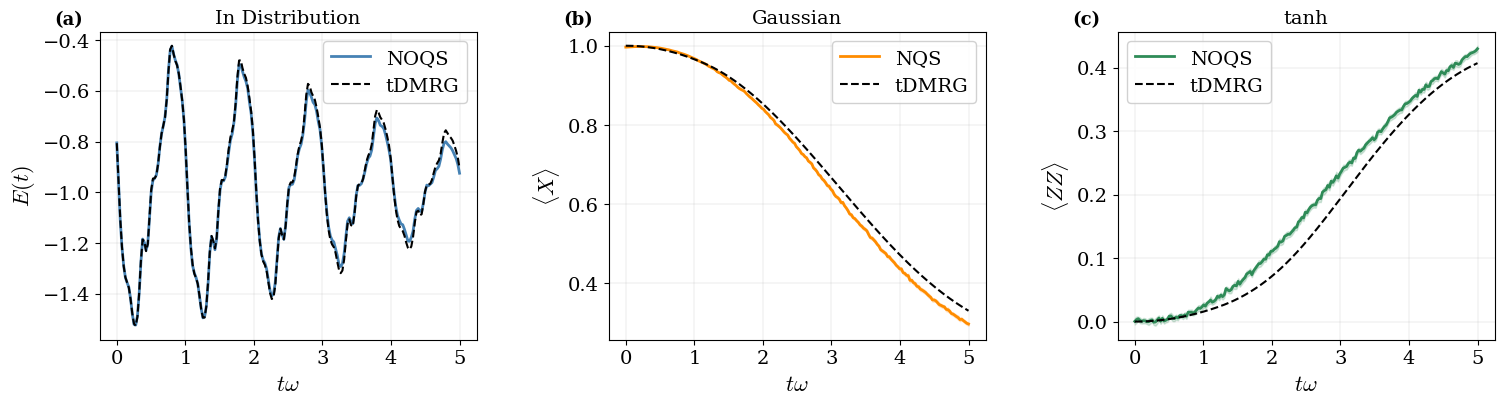}
    \caption{Performance of the NOQS on a $4 \times 8$ lattice ($N=32$), benchmarked against tDMRG ($\chi = 256$). The three driving protocols are the same as those in Fig.~\ref{fig:4by4}. \textbf{(a)} Energy $E(t)$ obtained from the NOQS predictions, matching almost perfectly with tDMRG results. \textbf{(b)}
    Average transverse magnetization $\left< X(t) \right>$ for the Gaussian pulse driving protocol. \textbf{(c)} $\langle ZZ(t)\rangle$ for a tanh ramp protocol. Both Gaussian and tanh protocols are out of the training distribution. Despite the exponentially larger Hilbert space compared to a $4 \times 4$ system, the NOQS predictions of local observables and correlators remain accurate.}
    \label{fig:4by8}
\end{figure*}

We stress that the NOQS framework holds for a given, fixed initial state. In this section, the initial state is chosen to be
\begin{equation}
    \ket{\psi_0} = \ket{+}^{\otimes N},
    \label{eq:plusinitstate}
\end{equation}
where $\ket{+}$ is the $+1$ eigenstate of $X_i$, and $N$ is the system size. In the computational basis, $\ket{\psi_0}$ corresponds to the uniform superposition over all $2^N$ states. The performance of our framework applied to another state, namely a ferromagnetically ordered initial state, will be discussed in Appendix.~\ref{app:ferromagnetic}.

To assess the performance of the wavefunction produced by the NOQS, we analyze three observables: the average transverse magnetization
\begin{equation}
    \langle X(t) \rangle = \frac{1}{N} \sum_i \bra{\psi_\theta(t)} X_i \ket{\psi_\theta(t)};
    \label{eq:magnetization}
\end{equation}
the nearest-neighbor $ZZ$ correlator:
\begin{equation}
    \langle ZZ(t) \rangle = \frac{1}{N_b} \sum_{\langle ij \rangle} \bra{\psi_\theta(t)} Z_i Z_j \ket{\psi_\theta(t)},
    \label{eq:zz_correlator}
\end{equation}
where $N_b$ is the number of nearest-neighbor bonds; and the energy 
\begin{equation}
    E(t) = \langle H(t) \rangle = \frac{1}{N} \bra{\psi_\theta(t)} H(t) \ket{\psi_\theta(t)},
    \label{eq:energy}
\end{equation}
where $H(t)$ is the time-dependent Hamiltonian in Eq.~\ref{eq:hamiltonian}.

The three observables probe complementary aspects of the time-evolved state $\ket{\psi(t)}$: $\langle X(t) \rangle$ is sensitive to the transverse polarization, $\langle ZZ(t) \rangle$ captures nearest-neighbor correlations, while $E(t)$ incorporates information from both the driving fields, as well as expectation values $\left<X\right>$, $\left<Z\right>$, $\left<ZZ\right>$. Comparing the accuracy for predicting the three observables tests the NOQS model's capability to capture not only the overall amplitude structure but also the spatial correlation buildup during time evolution.

We begin with a system of size $4 \times 4$, which can still be studied with Exact Diagonalization (ED) and provides a direct benchmark of our model's performance against numerically exact results. In Fig.~\ref{fig:4by4}(a), we consider an instance of driving fields $h_x(t)$ and $h_z(t)$ drawn from the family described by Eq.~\ref{eq:driving_field} that has \textit{not} been seen during training. The NOQS prediction of the energy $E(t)$ (Eq.~\ref{eq:energy}) matches almost perfectly with the exact results.

Remarkably, the NOQS also generalizes to driving protocols that are \textit{outside} the training distribution. In Fig.~\ref{fig:4by4}(b) and (c), we consider two experimentally realistic protocols: a Gaussian pulse and a ramp-up (represented by tanh) for the longitudinal and transverse fields.
Despite never encountering such functional forms during training, the NOQS demonstrates excellent predictive power and captures both $\left< X(t)\right>$ and $\left< ZZ(t) \right>$ accurately, confirming that the model has indeed learned a mapping between \textit{functional spaces}, instead of memorizing in-family trajectories.

Next we demonstrate the NOQS's performance on a $4 \times 8$ lattice ($N=32$ spins). Such system sizes are beyond the reach of exact numerical methods. This is indeed the regime in which NQS have been the most powerful, owing to the expressivity of neural networks that represent quantum states in an exponentially large Hilbert space. Here we benchmark the observables predicted by NOQS against time-dependent Density-Matrix Renormalization Group (tDMRG) calculations. We have verified convergence in the bond dimensions, and details about the implementation are in Appendix.~\ref{app:DMRG}.

In Fig.~\ref{fig:4by8}, we demonstrate the NOQS's performance for this larger system size. Across the three panels, we consider the same driving protocols as in Fig.~\ref{fig:4by4}. Despite the exponentially larger Hilbert space, the predicted observable dynamics tracks the results from tDMRG closely. These results highlight the scalability of our approach and its ability to capture dynamics in regimes where exact computations becomes prohibitive.

\subsection{Fine-Tuning}
While our NOQS model can be trained using physical loss alone, as detailed in Sec.~\ref{sec:training}, a key advantage of our framework is its ability to incorporate experimental data. In this section, we demonstrate this capability through fine-tuning.

In modern machine learning, foundation models such as large language models, are first \textit{pre-trained} on broad datasets, and then \textit{fine-tuned} on task-specific data to improve performance in targeted domains. Our training procedure in Sec.~\ref{sec:training} plays an analogous role: it serves as a general-purpose, foundational training stage. In this pre-training stage, the NOQS learns to map the function space of driving protocols to time-evolved quantum states. We can further fine-tune the model on a \emph{specific} instance of $h_x(t)$ and $h_z(t)$ using a small number of experimentally accessible measurements.

Concretely, suppose that for a particular driving protocol, measurements of $\langle X(t_m) \rangle$ and $\langle ZZ(t_m) \rangle$ are available at a \textit{sparse} set of points in time, $\{t_m\}$. Such exact results could be obtained from experiments or numerical simulations. We fine-tune the network parameters $\theta$ with a loss term that penalizes deviations of the NOQS predictions from these measured values:
\begin{align}
    \mathcal{L}_{\mathrm{data}} = \sum_m  &\left( \langle X(t_m) \rangle_\theta - \langle X(t_m) \rangle_{\mathrm{exp}} \right)^2 \nonumber \\
    &+ \left( \langle ZZ(t_m) \rangle_\theta - \langle ZZ(t_m) \rangle_{\mathrm{exp}} \right)^2 ,
    \label{eq:data_loss}
\end{align}
where the subscript $\theta$ denotes NOQS predictions and exp denotes the ground truths of $\left<X\right>$ and $\left< ZZ \right>$, possibly obtained from experiments.

\begin{figure}
    \centering
    \includegraphics[width=1\linewidth]{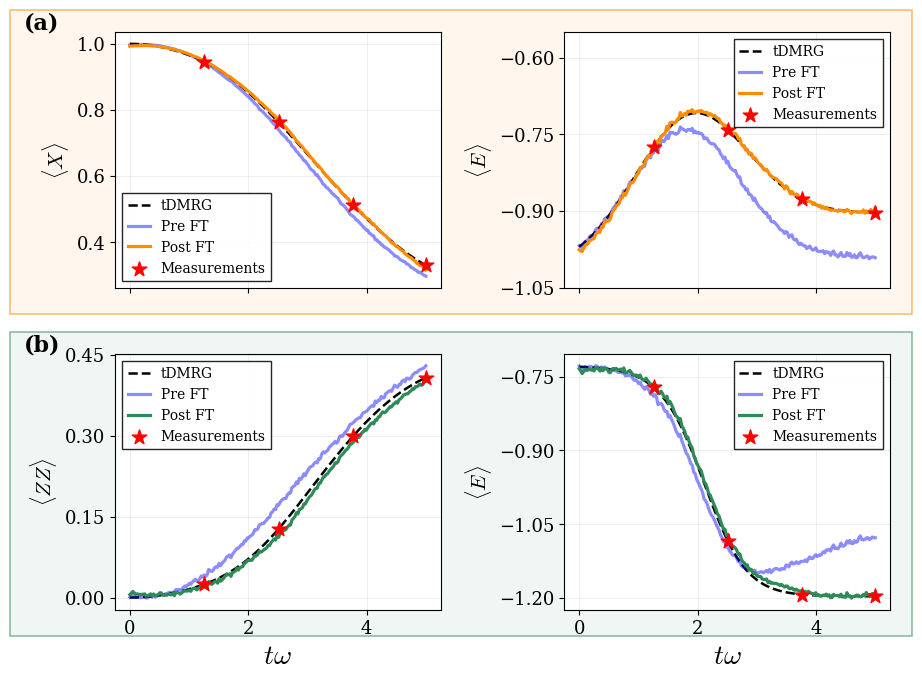}
    \caption{Performance of NOQS after fine-tuning, for \textbf{(a)} the Gaussian pulse and \textbf{(b)} tanh driving protocols, at a system size $4 \times 8$. Using only measurements of $X$ and $ZZ$ at four points in time, the Post Fine-Tune predictions become even more accurate for the out-of-distribution fields across the entire time interval.}
    \label{fig:fine_tuning}
\end{figure}

In Fig.~\ref{fig:fine_tuning}, we show the effect of fine-tuning for the two out-of-distribution protocols from Fig.~\ref{fig:4by8}: the Gaussian pulse in (b) and the tanh ramp in (c), for system size $4\times 8$. Using measurements of \textit{only} $\langle X \rangle$ and $\langle ZZ \rangle$ at four time points, the fine-tuned NOQS yields noticeably improved predictions for local observables and the energy, across the \textit{entire} time interval. 

We emphasize that in contrast to Ref.~\cite{nqs-finetuning}, which trains NQS further on unseen system parameters, our fine-tuning stage only involves the data loss (Eq.~\ref{eq:data_loss}), and the TDVP loss is no longer used. The improvements in $E(t)$, which involves not only $\left<X\right>$ and $\left<ZZ\right>$, but also $\left< Z \right>$, reflect the fact that fine-tuning on sparse measurements improves the quality of the full underlying wavefunction. This is possible because the pre-trained model already encodes a physically consistent state; the sparse data serve to further refine it, rather than reconstruct it from scratch.

Our result highlights a practical advantage of the NOQS framework: a pre-trained model can already predict local observables accurately, for unseen instances of driving (Fig.~\ref{fig:4by8}). With only a handful of easily accessible measurements, one can increase the accuracy even further for specific driving protocols. Our framework can be readily adapted to specific experimental conditions, without retraining from scratch. Moreover, the fine-tuning stage is extremely cheap, as the loss function is simply trying to penalize deviations from the sparse measurements (Eq.~\ref{eq:data_loss}). The workflow in our model mirrors the pre-train and fine-tune paradigm that has proven effective across machine learning, and it opens the door to hybrid computational-experimental approaches for quantum dynamics.

\subsection{Transferability Across Discretizations \label{sec:discretization_invariance}}
Compared to conventional neural networks, Fourier Neural Operators (FNO) have a remarkable capacity to transfer between different discretizations of the underlying domain~\cite{FNO_Li, NeuralOperator_general}. This property arises because the FNO parameterizes the integral kernel in frequency space, rather than learning pointwise mappings tied to specific grids. The Fourier representation can then be evaluated at arbitrary time points, regardless of the underlying resolution. For classical partial differential equations, discretization invariance has been established as a key advantage of FNO-based solvers, enabling zero-shot generalization to finer spatiotemporal grids without retraining~\cite{FNO_Li}. This transferability has also been demonstrated for FNO treatments of time-periodic quantum systems~\cite{FNOFloquet}. 

In our current architecture, the time-dependence of the quantum state comes \textit{entirely} from the context tokens $M(t)$, which are produced by the FNO. This architectural choice of NOQS allows it to inherit the temporal discretization invariance of the FNO.

\begin{figure}[t]
    \centering
    \includegraphics[width=1\linewidth]{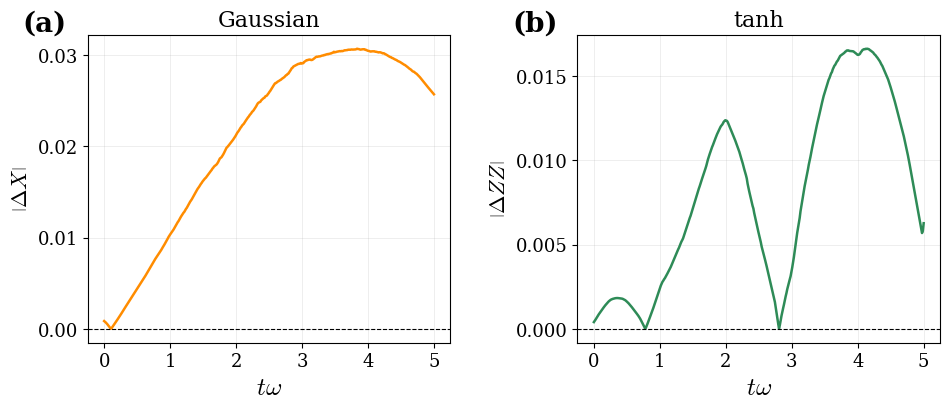}
    \caption{Temporal super-resolution of the NOQS on a $4 \times 4$ lattice. The model is trained on $N_t = 200$ time points and evaluated on $N_t = 400$ points, \textit{without retraining}. \textbf{(a)} Absolute error $|\Delta X(t)| = |\langle X(t)\rangle_{\mathrm{NOQS}} - \langle X(t)\rangle_{\mathrm{exact}}|$ for a Gaussian pulse driving protocol. \textbf{(b)} Absolute error $|\Delta ZZ(t)|$ for a tanh ramp protocol. The smooth error profiles confirm that the NOQS inherits the discretization invariance of the Fourier Neural Operator. Trained on a coarse grid, the NOQS is capable of predicting at finer temporal resolutions accurately.}
    \label{fig:super_resolution}
\end{figure}

To test this, we take the NOQS that has been trained on a temporal grid of $N_t = 200$ evenly spaced time points. Without any retraining, fine-tuning, or interpolation, we evaluate the model on a finer grid of $N_t = 400$ points. This constitutes a zero-shot temporal super-resolution: the model is asked to predict at times it has never encountered during training. In Fig.~\ref{fig:super_resolution}, we show the absolute deviation of the NOQS predictions from exact computations for $\langle X(t) \rangle$ under a Gaussian pulse protocol and $\langle ZZ(t) \rangle$ under a tanh ramp. The error profiles are smooth and remain small throughout the time interval, with no artifacts, discontinuities, or oscillations at the scale of the training grid spacing. This confirms that the full NOQS architecture exhibits discretization invariance.
 
This property furthers the practical value of our framework. An experimentalist interested in resolving fast dynamics, or in comparing predictions against data acquired at a non-uniform set of measurement times, can query the same pre-trained model on other temporal grids. In contrast, methods that discretize time evolution into fixed steps, such as Trotterized circuits or finite-difference TDVP integrators, inevitably require either retraining or additional computational steps to increase the temporal resolution. Thus, the NOQS architecture, by design, combines the flexibility of a continuous-time representation with the expressivity of a discrete many-body ansatz.
 
\section{Conclusion and Outlook \label{sec:conclusion}}
In this work, we propose the Neural Operator Quantum State (NOQS), a novel architecture that learns the solution operator of the Schr\"odinger equation across a function space of driving protocols. By coupling a neural operator realized via FNO, which processes the continuous, temporal structure of the driving fields, to a transformer-based autoregressive wavefunction through cross-attention, the NOQS predicts time-evolved quantum states not only for unseen protocols in distribution, but also for out-of-distribution but experimentally relevant driving fields. The entire model is trained in a self-supervised way using a physically motivated loss, requiring no external data from exact numerics, tensor networks, or experiments. 
 
As a numerical demonstration, we applied the NOQS to study a two-dimensional transverse-field Ising model with time-dependent transverse and integrability-breaking longitudinal fields. Our model accurately predicts local observables and correlators, which are benchmarked against exact diagonalization and time-dependent DMRG. Crucially, the NOQS generalizes to out-of-distribution protocols, including Gaussian pulses and ramps, which are qualitatively different from the truncated Fourier series used during training. We further demonstrated that the NOQS inherits the discretization invariance of its FNO backbone: a model trained on $N_t = 200$ time points produces accurate predictions when evaluated on a finer grid of $N_t = 400$ points, achieving zero-shot temporal super-resolution. Both results confirm that the model learns a genuine \textit{functional} mapping rather than memorizing individual trajectories on fixed discretizations. Finally, we showed that a pre-trained NOQS can be fine-tuned with sparse measurements of local observables, improving the accuracy of the full quantum state, at negligible additional cost.

There are a few natural directions for future work. First, the current NOQS can transfer across a function space of protocols, but at fixed coupling constants (namely $J$). An architecture that additionally conditions on static parameters such as interaction strengths, disorders, or dissipation rates would enable a single model to span a joint space of protocols and static parameters. This direction could be relevant for driven-dissipative quantum systems~\cite{driven_dissipative1, driven_dissipative2, driven_dissipative3}, which depend on both time-dependent functions, as well as dissipation strengths.

On the more theory side of investigations, one could ask: what types of driving fields are easier or harder to learn? Does the learnability transition in the presence of disorder or near critical points? Finally, how much information does the context token $M(t)$ encode? Can we learn some aspects of the quantum dynamics from the evolution of these context tokens? Insights into these questions will allow for a better understanding of the limitations and capabilities of the NOQS model.

\begin{acknowledgments}
ZQ acknowledges discussions with Junkai Dong.
YP is supported by the US National Science Foundation (NSF) Grants  No.\ PHY-2216774 and No.\ DMR-2406524.
\end{acknowledgments}

\appendix
\section{Hyperparameter Choice and Loss Implementation~\label{app:hyperparameters}}
To ensure reproducibility, in this Appendix, we list the details of our model architecture and training procedure.

As discussed in Sec.~\ref{sec:architecture} in detail, our model architecture consists of a Fourier Neural Operator (FNO) and a transformer, which are coupled through a cross-attention mechanism. Parameters in the Transformer consist of the number of decoder layers, the latent (embedding) space dimension, as well as the number of attention heads. After the decoder blocks, there is also a two-layer, fully-connected unembedding network with intermediate dimension $d_f$. Similarly, parameters of the FNO include the number of layers, the model width, and the number of modes that are kept in frequency domain. We have empirically verified that the model performance depends most strongly on the FNO width and transformer embedding dimension. We list the hyperparameters for the NOQS model in the first two parts of Table.~\ref{tab:hyperparameters}.

We also list the hyperparameters used in training and fine-tuning. We use the Adam optimizer with a scheduled, decaying learning rate, which helps stabilize training. Other details about training are included in Table.~\ref{tab:hyperparameters}.

In practice, rather than minimizing the raw expectation $\mathcal{L}_{\mathrm{TDVP}}(t)$ in Eq.~\eqref{eq:tdvp_local} directly, we minimize the variance of $|L_{\mathrm{loc}}(\bm{\sigma}, t)|^2$ over the sample configurations $\bm{\sigma}$ at each time $t$. This choice is motivated by the following argument: if the optimization is exact, so that $L_{\mathrm{loc}}(\bm{\sigma}, t)$ becomes $\bm{\sigma}$-independent, i.e.,
\begin{equation}
    L_{\mathrm{loc}}(\bm{\sigma}, t) \equiv c(t), \quad \forall\, \bm{\sigma},
\end{equation}
for some scalar $c(t) \in \mathbb{R}$, then the Schr\"{o}dinger residual reduces to a global, configuration-independent phase. Since physical states are defined only up to a global phase, such a solution corresponds to the correct physical state satisfying the time-dependent Schr\"{o}dinger equation. We therefore minimize the \textit{variance}
\begin{equation}
    \mathrm{Var}_{\bm{\sigma}}\!\left[L_{\mathrm{loc}}(\bm{\sigma}, t)\right] 
    = \mathbb{E}_{\bm{\sigma}}\!\left[\left|L_{\mathrm{loc}}(\bm{\sigma}, t) 
    - \overline{L_{\mathrm{loc}}(t)}\right|^2\right],
\end{equation}
where $\overline{L_{\mathrm{loc}}(t)}$ denotes the sample mean. This serves as a more stable and gauge-invariant training objective: it drives the residual toward a $\bm{\sigma}$-independent constant, thereby enforcing the correct quantum dynamics while remaining insensitive to the unphysical global phase.

\begin{table}[ht]
\centering
\label{tab:hyperparameters}
\begin{tabular}{ll}
\hline
\textbf{} & \textbf{Hyperparameter} \\ \hline
\textbf{Architecture (Transformer)} & \\
Number of Decoder Layers ($N_T$) & 3 \\
Embedding Dimension ($d_e$) & 128 \\
Number of Attention Heads ($n_h$) & 8 \\
Feed-forward Dimension ($d_f$) & $4 \times d_e = 512$ \\
Activation Function $\sigma$ & GeLU~\cite{gelu} \\
\hline
\textbf{Architecture (FNO)} & \\
Number of FNO Layers ($N_F$) & 3 \\
FNO Width ($d_v$) & 96 \\
FNO frequency modes ($k_\text{max}$) & 64 \\
Number of Context Tokens ($N_C$) & 4 \\
Number of Points in Time & 200\\
\hline
\textbf{Training} & \\
Optimizer & Adam~\cite{adam} \\
Initial Learning Rate (LR) & $4 \times 10^{-4}$ \\
LR decay factor & 0.95 \\
LR decay rate & 2000 (steps) \\
Minimum LR & $4 \times 10^{-6}$ \\
Batch size ($B$) & 4 \\
Time points per step $(K)$ & 3 \\
MC sample per step $(M)$ & 128 \\
Training Steps & 60000 \\
Weight of Anchor Loss ($\lambda_w$) & 10.0 \\
\hline
\textbf{Fine Tuning} & \\
Training Steps & 300 \\
Learning Rate & $3 \times 10^{-4}$ \\
Number of Data Points & 4 \\
\hline

\end{tabular}
\caption{Hyperparameters for the Neural-Operator Quantum State Model, for system size $4 \times 4$. For the system size $4 \times 8$, the only changes are in the number of decoders, FNO layers, and context tokens: $N_T=3\rightarrow 4$; $N_F = 3\rightarrow 4$; $N_C =4 \rightarrow 8$.}
\end{table}

\section{Technical Details on tDMRG Implementation \label{app:DMRG}}
For a system of size $4 \times 8$, exact time evolution is no longer feasible due to the exponential growth of system size. We therefore use time-dependent Density Matrix Renormalization Group (tDMRG) as a benchmark for observable dynamics. 

We choose the threshold for SVD truncation to be $10^{-12}$. The bond dimension $\chi$ has been verified to converge; we fix $\chi = 256$ in Fig.~\ref{fig:4by8} throughout for comparison with NOQS predictions. The tDMRG calculations are implemented with the TeNPy package~\cite{tenpy, tenpy2024}.

\section{NOQS Performance on a Ferromagnetic Initial State \label{app:ferromagnetic}}
In the main text, we have focused on the case with a paramagnetically ordered initial state (Eq.~\ref{eq:plusinitstate}). However, our NOQS architecture as well as training procedure can be readily adapted to different initial state. In this Appendix, we demonstrate the model's performance for the ferromagnetically ordered initial state:
\begin{equation}
    \ket{\psi_0^\text{ferro}} = \bigotimes_i\ket{\uparrow}_i,
    \label{eq:ferromagnetic_state}
\end{equation}
where $\ket{\uparrow}_i$ denotes the eigenstate of $2Z_i$ with eigenvalue $+1$. 

\begin{figure}[!htbp]
    \centering
    \includegraphics[width=1\linewidth]{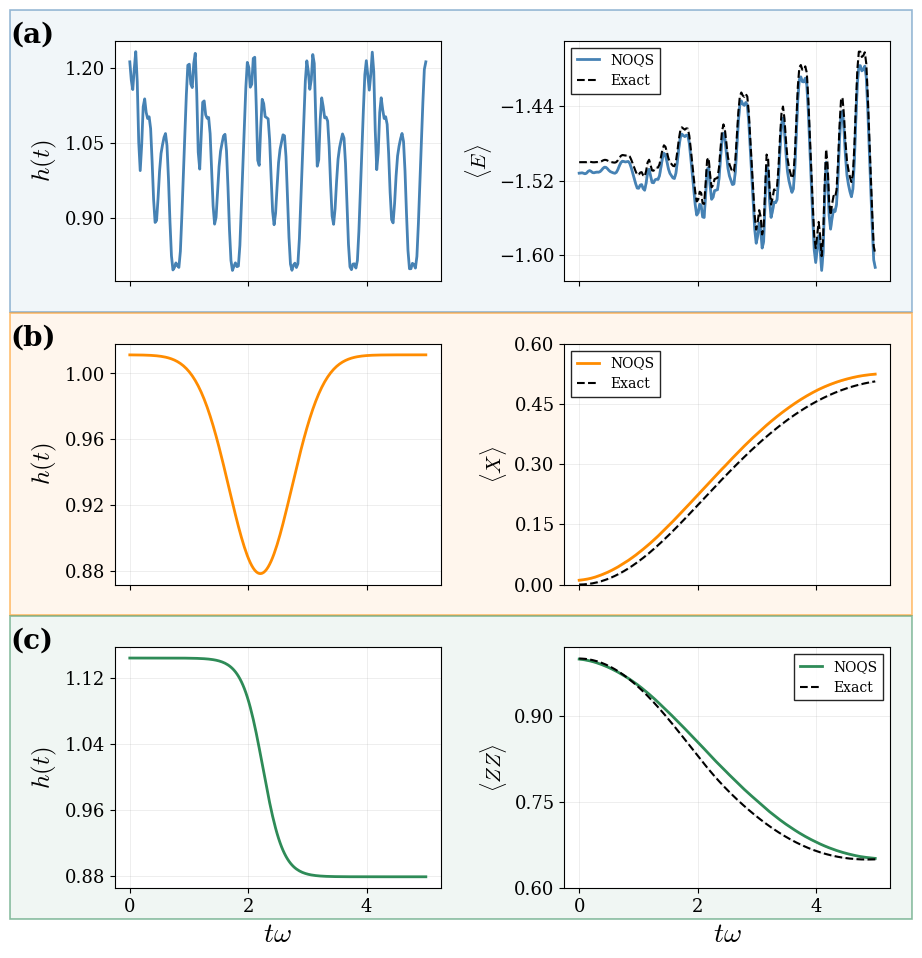}
    \caption{NOQS performance on a ferromagnetically ordered initial state (Eq.~\ref{eq:ferromagnetic_state}) benchmarked against exact results, for a $4 \times 4$ lattice. For both \textbf{(a)} in-distribution and \textbf{(b,c)} out-of-distribution driving protocols, the NOQS accurately captures the local observables. }
    \label{fig:ferromagnetic}
\end{figure}

In Fig.~\ref{fig:ferromagnetic}, we demonstrate that for the ferromagnetic initial state, the NOQS captures local observables accurately, again for both in-distribution and out-of-distribution driving protocols. We highlight that the drivings are different from those in the main text (Fig.~\ref{fig:4by4}), which further supports that the NOQS achieves learning not for particular protocol instances, but over a function space of drivings.

\bibliography{main}

\end{document}